\documentclass{iopart}
\usepackage{graphicx}
\usepackage{amssymb}

\renewcommand{\Tr}{{\rm Tr}}

\newcommand{\be}{\begin{equation}}
\newcommand{\ee}{\end{equation}}
\newcommand{\tn}{\widetilde{N}}

\begin{document}

\title{Exponentially small quantum correction to conductance}
\author{Lucas H. Oliveira$^1$, Pedro H. S. Bento$^2$Marcel Novaes, $^1$}
\address{$^1$ Instituto de F\'isica, Universidade Federal de Uberl\^andia, 38408-100, Brazil}
\address{$^3$ Instituto de F\'isica, Universidade Federal de Goiás, 74690-900, Brazil}

\date{\today}

\begin{abstract}

When time-reversal symmetry is broken, the average conductance through a chaotic cavity, from an entrance lead with $N_1$ open channels to an exit lead with $N_2$ open channels, is given by $N_1N_2/M$, where $M=N_1+N_2$. We show that, when tunnel barriers of reflectivity $\gamma$ are placed on the leads, two correction terms appear in the average conductance, and that one of them is proportional to $\gamma^{M}$. Since $M\sim \hbar^{-1}$, this correction is exponentially small in the semiclassical limit. Surprisingly, we derive this term from a semiclassical approximation, generally expected to give only leading orders in powers of $\hbar$. Even though the theory is built perturbatively both in $\gamma$ and in $1/M$, the final result is exact.

\end{abstract}

\maketitle

\section{Introduction}

The semiclassical approximation has always been, from the very beginnings of quantum mechanics, a way to make sense of the wave nature of matter in the microscopic scale, and has been developed ever since, both in depth and in scope \cite{book1,book2,heller,berry}. The fascinating problem of quantum tunneling, in particular, has always challenged our understanding and been the subject of semiclassical investigation \cite{book3,book4}. 

Generally speaking, a semiclassical approach to a quantum problem is expected to be effective when Planck's constant, $\hbar$, is much smaller than all other actions present, and should work perturbatively, i.e. by taking into account leading orders in Taylor expansions of the form $c_0+c_1\hbar+c_2\hbar^2+\cdots$. However, in some cases it is possible to go 'beyond all orders' and recover quantum effects that contain factors of the form $e^{-c/\hbar}$. This includes level splitting in one-dimensional symmetric quantum wells \cite{holstein,garg} and one-dimensional propagation of localized wavepackets \cite{hagedorn1, hagedorn2, toloza}.

One might argue that treating a single dimension is an oversimplification. Could exponential accuracy be achieved by the semiclassical approximation in a more general problem, like a chaotic system? We show in the present work that the answer is actually positive. 

We consider quantum transport through ballistic systems with chaotic dynamics, like a two dimensional electron gas confined in mesoscopic cavity which is attached to the outside world by infinite leads \cite{nazarov}. This is a more sophisticated version of the tunnelling problem that has long been a prime testing ground for ideas from quantum chaos, the study of the interplay between unpredictability due to dynamics and quantum uncertainty \cite{qc1,qc2,qc3,qc4}. One of the main findings in this area has been universality: average observables are insenstive to system's details and depend only on the symmetries that are present \cite{haake}.

We consider two leads, supporting $N_1$ and $N_2$ open channels, as sketched in Fig. 1, and spinless particles with broken time-reversal symmetry, i.e. in the so-called unitary class. Quantum scattering in this context is described by the $M\times M$ unitary $S$ matrix relating incoming to outgoing quantum amplitudes, where $M=N_1+N_2$. Let $t$ be the block from $S$ which contains the transmission amplitudes between the leads. Then the Hermitian matrix $t^\dagger t$ encodes the relevant transport properties that characterize the electrical current as a funcion of time and energy \cite{landauer,buttiker}. The (dimensionless) conductance of the system is given by $g=\Tr(t^\dagger t)=\sum_{io}|t_{io}|^2$. It is a widely fluctuating function of the energy and one considers its average value $\langle g\rangle$, within a range of energies which is classically small but large in the quantum scale. 

When the leads connecting the cavity to the outside world are ideal, i.e. perfectly transmitting, it is well known that the transmission probability between any pair of channels is $1/M$ and the average conductance is $N_1N_2/M$. A more realistic setting is to assume the presence of tunnel barriers in the leads \cite{bar1,bar2,bar3,bar4}, so that channel $i$ has an associated tunnelling rate $\Gamma_i$, with $\Gamma_i = 1$ being the ideal case. We assume for simplicity that all such rates are equal and introduce the reflectivity $\gamma=1-\Gamma$. 

\begin{figure}[t]
\includegraphics[scale=1]{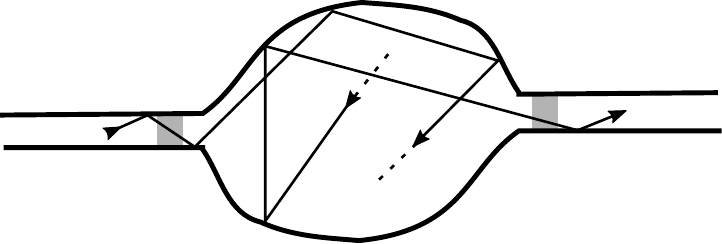}
\caption{Schematic representation of the system. The electron travels along the first waveguide ($N_1$ open channels), tunnels through a barrier and enters a cavity, where its dynamics is chaotic (the convoluted nature of the typical trajectory, in between the dashed lines, is not shown for simplicity). It then tunnels out to the second waveguide ($N_2$ open channels). The tunnel barriers are represented by the grey rectangles.}
\end{figure}

Brouwer and Beenakker considered the conductance problem in the presence of barriers within a random matrix theory formulation \cite{rmt}, which forgoes dynamical details and treats the $S$ matrix statistically \cite{BB2}. In particular, they obtained the distribution of $g$ for a single open channel in each lead and showed that, in that case, $\langle g\rangle = 1/2-2\gamma/3+\gamma^2/6$ (their actual expression is a bit cumbersome, but equivalent to this). They also showed \cite{BB} that, to leading order for large $M$, we have $\langle g\rangle =\tn_1\tn_2/\widetilde{M}$, where $\tn_i=\Gamma N_i$ and 
\be \widetilde{M}=\Gamma M=(1-\gamma)M\ee
work like effetive numbers of channels.

The objective of the present work is to present a semiclassical calculation of $\langle g\rangle$ that is valid for arbitrary values of $M$. We originally expected to find a quadratic function of $\gamma$, with simple $M$-dependent coefficients. To our surprise, we found that this is not the case: in the general result the quadratic term is replaced by something proportional to $\gamma^M$. 

Quantum corrections to the macroscopic $N_1N_2/M$ conductance are usually called 'weak localization' effects \cite{qc2,weak2}. On the other hand, asymptotic approximations that go beyond all orders are sometimes called 'superasymptotics' \cite{super1,super2}. We therefore call our result the 'superweak correction'.

What we show is that
\be\label{gu} \langle g\rangle=\frac{N_1N_2}{M}-\frac{N_1N_2M\gamma}{M^2-1}+\frac{N_1N_2\gamma^M}{M(M^2-1)}.\ee The first two terms are macroscopic, i.e. of order $M$ for large $M$. The last term, however, the superweak correction, decreases exponentially with $M$ for $0<\gamma<1$. Since $M\sim \hbar^{-1}$, this non-perturbative term is of the form $e^{-c/\hbar}$. It is missing from all previous approaches to the problem \cite{whitney, jacquod, kuipersrichter} (Kuipers \cite{kuipers}, for example, finds evidence only of the first two terms in his perturbative approach that includes several leading orders in powers of $1/M$), including ones based on random matrices \cite{kanz1,vidal,kanz2,perez}. 

This correction should be contrasted with the usual weak localization effect, which is of order $1/M\sim \hbar$, and with the correction due to a finite Ehrenfest time $\tau_E$ obtained in \cite{Ehren1,Ehren2}. The latter is proportional to $e^{-\tau_E/\tau_D}$, where $\tau_D$ is the classical dwell time of the cavity; since $\tau_E\sim \log(\hbar)$, this correction is also of order $\hbar$.

Notice that even though the superweak correction is exponentially small for large $M$, it is in fact essential for the result to be physically meaningful. For instance, if this term is omitted the resulting expression for $\langle g\rangle$ would become negative for $\gamma>1-1/M^2$. In particular, when $N_1=N_2=1$ we have $\langle g\rangle=1/2-2\gamma/3+\gamma^2/6$, in agreement with the corresponding random matrix theory result \cite{BB2}. 

\section{Semiclassical formulation} 

We employ the trajectory based semiclassical approximation, with the inclusion of subtle action correlations that were introduced by Richter and Sieber \cite{sieber1,sieber2} and further refined by Haake and collaborators \cite{essen3,essen4,essen5}. The matrix element $t_{oi}$ is written in terms of scattering trajectories entering the system through channel $i$ and exiting through channel $o$. When correlations among scattering trajectories are taken into account, and the required integrations over phase space are performed, the theory has a diagrammatic formulation. This is in principle a perturbative theory in the parameter $M^{-1}$. We refer the reader to previous works \cite{essen3,essen5,greg1,greg2} for details. 

As discussed in \cite{whitney,kuipers}, when tunnel barriers are present the semiclassical diagrammatic rules to compute the transmission probability are as follows: each edge in a diagram corresponds to multiplication by $\widetilde{M}^{-1}$; each vertex of valence $2q$ corresponds to multiplication by $-M(1-\gamma^q)$; each trajectory that enters of leaves the cavity produces a factor $(1-\gamma)$, to account for transmission through the barriers; finally, a factor of $\gamma$ arises each time a trajectory experiences an internal reflection at a lead. 

For example, two of the simplest diagrams that contribute to the semiclassical calculation of conductance are shown in Fig. 2. They have been discussed before in detail \cite{essen3,kuipers,pedro}. The one on panel a) has four edges, one vertex of valence six and no internal reflections; its total contribution is 
\be -(1-\gamma)^2\frac{M(1-\gamma^3)}{\widetilde{M}^4}.\ee
The one on panel b) has four edges, no vertices but two internal reflections; its total contribution is 
\be (1-\gamma)^2\gamma^2\frac{1}{\widetilde{M}^4}.\ee

\begin{figure}[t]
\includegraphics[scale=0.6]{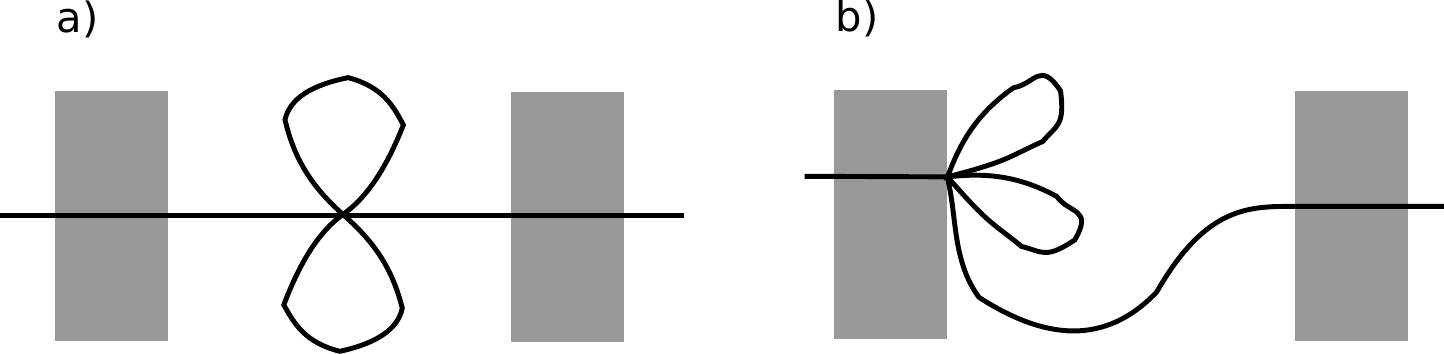}
\caption{Two diagrams from the semiclassical calculation of conductance. The tunnel barriers are represented by the grey rectangles.}
\end{figure}

As discussed in \cite{pedro}, the diagrammatic rules that apply to systems with tunnel barriers can be encoded, when computing the average conductance, by means of a matrix intgral. This integral is designed so that, when it is evaluated using ribbon graph techniques derived from Wick's rule, the summation over diagrams is equivalent to the semiclassical theory. In this way, all diagrams are taken into account automatically, to all orders in perturbation theory, providing results that are exact in the parameter $M$. This efficient semiclassical approach was introduced in \cite{matrix} and later succesfully applied to a range of different problems \cite{trs,corr,sebas}. 

In the present case, the adequate integral is given as
\be\label{comp} \langle |t_{io}|^2 \rangle=\lim_{N\to 0}(1-\gamma)^2\frac{1}{\mathcal{Z}}\int \mathcal{I}(Z,Z^\dagger)\mathcal{C}(Z,Z^\dagger)dZ.\ee
The integrand has two components. First,
\be\mathcal{I}(Z)=e^{-\sum_{q=1}^\infty\frac{1}{q}(M(1-\gamma^q)\Tr(Z^\dag Z)^q} \ee
is called the internal part, containing information about what happens inside the chaotic scattering region. When the exponential is Taylor expanded, it generates contributions corresponding to all possible vertices and edges. Second,
\be \mathcal{C}(Z,Z^\dagger)= \left(\frac{1}{1-\gamma Z^\dag Z}Z^\dagger\right)_{io}\left(Z\frac{1}{1-\gamma Z^\dag Z}\right)_{oi}\ee
is called the channel part, responsible to coupling with with leads at incoming channel $i$ and outgoing channel $o$. The geometric series generates contributions corresponding to all possible internal reflections. Finally, $\mathcal{Z}=\int e^{-\widetilde{M}\Tr(Z^\dag Z)}dZ$ is a normalization constant.

After the integration over $Z$ has been performed, the limit $N\to 0$ must be taken \cite{matrix}. This is important to remove spurious contributions which, semiclassically, correspond to periodic orbits trapped in the open cavity and which do not contribute to conductance.

\section{The integrals} 

The matrix integral is computed by means of the singular value decomposition $Z=UDV^\dagger$, with $U$ and $V$ uniformly distributed over the unitary group (taken to have unit volume). The Jacobian of this decomposition is $|\Delta(X)|^2$ \cite{jacob}, where $X=D^2$ is a diagonal matrix with the same eigenvalues, $\{x_1,\ldots,x_N\}$, of $Z^\dagger Z$ and 
\be \Delta(X)=\prod_{1\le i<j\le N}(x_j-x_i) \ee
is the so-called Vandermonde.

This leads to 
\be \mathcal{Z}=\int_0^\infty e^{-\widetilde{M}{\rm Tr}(X)}|\Delta(X)|^{2}dX,\ee which is a Selberg integral \cite{selberg}, known to be equal to \be \mathcal{Z}=\widetilde{M}^{-N^2}\prod_{j=1}^{N}(j-1)!j!.\ee

The internal part is independent of $U,V$. The channel part is 
\be \sum_{a,b=1}^N V_{ia}\frac{D_a}{1-\gamma_2 X_a}U^\dagger_{ao}U_{ob}\frac{D_b}{1-\gamma_1 X_b}V^\dagger_{bi}.\ee Using the orthogonality relations of matrix elements of the unitary group,\be\int U_{ab}U^\dagger_{cd}V_{ef}V^\dagger_{gh}dUdV=\frac{1}{N^2}\delta_{ad}\delta_{bc}\delta_{eh}\delta_{fg}\ee the result of the angular integral of the channel part is simply 
\be \label{C} \widehat{\mathcal{C}}(X)= \frac{1}{N^2} {\rm Tr}\left(\frac{X}{(1-\gamma X)^2}\right).\ee
The sums over $i,o$ produce a factor $N_1N_2$ which, combined with the prefactors $(1-\gamma)^2$ already present, becomes $\tn_1\tn_2$.

We are left to deal with the eigenvalue integral
\be\label{E} \langle g \rangle=\tn_1\tn_2\lim_{N\to 0}\frac{1}{\mathcal{Z}}\int _0^1|\Delta(X)|^{2}
\mathcal{I}(X)\widehat{\mathcal{C}}(X)dX,\ee
with 
\be \mathcal{I}(X)= \frac{\det(1-X)^{M}}{\det(1-\gamma X)^M},\ee
where we have used 
\be e^{-N\sum_{q=1}^\infty \frac{1}{q} {\rm Tr}X^q}=e^{N{\rm Tr}\log(1-X)}=\det(1-X)^{N}.\ee

In the following calculations we rely on some well known results about Schur polynomials  $s_\lambda(X)$ \cite{macdonald}, which are homogeneous symmetric polynomials in the eigenvalues of $X$, labeled by an integer partition $\lambda=(\lambda_1,\ldots,\lambda_\ell)$. For example, the Cauchy expansion 
\be \det(1-\gamma X)^{-M}=\sum_{\lambda}\gamma^{|\lambda|}s_\lambda(X)s_\lambda(1^M),\ee where $1^M$ is a $M$-dimensional identity matrix and $|\lambda|=\sum_i \lambda_i$, can be used with the internal part. The channel part,
\be \widehat{\mathcal{C}}(X)=\sum_{n=0}^\infty(n+1)\gamma^{n}{\rm Tr}(X^{n}),\ee
can also be expanded in terms of Schur polynomials,
\be \widehat{\mathcal{C}}(X)=\sum_{n=0}^\infty(n+1)\gamma^{n}\sum_{\mu\vdash n}\chi_\mu(n) s_\mu(X),\ee where $\chi_\mu(\lambda)$ are the characters of the irreducible representations of the permutation group \cite{macdonald}. It is well known that $\chi_\mu(n)=0$ unless $\mu$ is a so-called hook partition, $\mu=(n-k,1^k)$, in which case it equals $(-1)^k$.

There are two Schur polynomials expansions in our integrand. We must therefore bring into play the Littlewood-Richardson coefficients, defined as 
\be\label{LR1} s_\lambda(X) s_\mu(X)=\sum_{\theta\vdash |\lambda|+|\mu|} C_{\lambda,\mu}^\theta s_\theta(X).\ee These coefficients can then be contracted with $s_\lambda(1^M)$ to produce a skew Schur polynomial, \cite{macdonald}
\be \sum_{\lambda}\gamma^{|\lambda|}C_{\lambda,\mu}^\theta s_\lambda(1^M)=\gamma^{|\theta|-n}s_{\theta/\mu}(1^M).\ee

The integral that must be done is
\be \frac{1}{\mathcal{Z}}\int _0^1|\Delta(X)|^{2}\det(1-X)^{M}s_\theta(X)dX,\ee
a generalization of the Selberg integral \cite{kaneko,kadell} which is given by \be \widetilde{M}^{N^2}\left([N]_\theta\right)^2\frac{d_\theta}{|\theta|!}\prod_{j=1}^N\frac{(M+N-j)!}{(\theta_j+M+2N-j)!}.\ee Here $d_\theta=\chi_\theta(1)$ is the dimension of an irreducible representation of the permutation group and 
\be [N]_\theta=\prod_{j=1}^{N}\frac{(\lambda_j+N-j)!}{(N-j)!}.\ee

\section{Taking $N\to 0$} 

Having computed all the integrals, we must consider the limit $N\to 0$. First of all,
\be \prod_{j=1}^N\frac{(M+N-j)!}{(\theta_j+M+2N-j)!}\to\frac{1}{[M]_\theta}.\ee
Also, we have that, for small $N$, \cite{macdonald} 
\be ([N]_\theta)^2= \frac{(|\theta|-1)!^2}{d_\theta^2} N^{2D(\theta)}+O(N^{2D(\theta)+2}),\ee 
where $D(\theta)$ is the size of the Durfee square of the partition $\theta$. Because of the $N^2$ in the denominator of Eq.(\ref{C}), the limit requires that $D(\theta)=1$, which means that $\theta$ must be a hook. If we write $\theta=(n-k+r,1^{k+q})$, then 
\be d_\theta= \frac{(n+r+q-1)!}{(n-k+r-1)!(k+q)!},\ee
and $[M]_{\theta} = (M-k-q)^{(n+r+q)}$ is a rising factorial

Moreover, because $\mu=(n-k,1^k)$ is also a hook, the Young diagram of the skew partition $\theta/\mu$ has two disjoint pieces, $\theta/\mu=(r)\cup (1^q)$. Then the skew polynomial factors as a product of rising and falling factorials, $s_{\theta/\mu}(1^M) = (M)^{(r)}(M)_{(q)}/r!q!.$ We therefore replace the sum over $\theta$ with sums over $r$ and $q$.
This leads to
\begin{eqnarray}
\fl   \langle g\rangle = \tn_1\tn_2\sum_{n,r,q\ge 0}\sum_{k=0}^{n-1}\frac{(-1)^k(n-k+r-1)!(k+q)!}{r!q!(n+r+q)}\frac{(n+1)(M)^{(r)}(M)_{(q)}\gamma^{n+r+q}}{(M-k-q)^{(n+r+q)}}.
\end{eqnarray}
Changing variables and orders of summation, and using that
\be\fl \sum_{n=k}^{l-r} \frac{(-1)^n(n+1)(M)_{(l-n-r)}}{(l-n-r)!}=(-1)^{k}\frac{(Mk+M-l+r-1)}{(l-r-k)!}\frac{(M-2)!}{(M+k+r-l-1)!},\ee
we obtain  
\be\label{gf} \fl
\langle g\rangle = \tn_1\tn_2(M-2)!\sum_{l=0}^\infty \frac{\gamma^l}{(l+1)} \sum_{r=0}^{l}\sum_{k=0}^{l-r}\frac{(k+r)!}{r!}\frac{(M)^{(r)}}{(M+k+r)!}(Mk+M-l+r-1).
\ee

The sums over $r,k$ can be computed to give
\be
\langle g\rangle = \tn_1\tn_2\frac{(M-2)!}{(M+1)!}\sum_{l=0}^\infty \gamma^l(M^2-l-1),
\ee
a simple geometric series which cancels with the $(1-\gamma)^2$ from $\tn_1\tn_2$, leading to precisely the first two terms in the conductance expression, Eq.(\ref{gu}). We remark that, even though the theory has been built perturbatively both in $\gamma$ and in $1/M$, it leads to an exact result.

However, the superweak correction is still missing. In order to obtain it, we must perform a kind of regularization, replacing $(M+k+r)!$ by $(M+\epsilon+k+r)!$ in the denominator of Eq.(\ref{gf}). Then, the term of order $\gamma^s$, with $s>1$, is equivalent, when $\epsilon\to 0$, to
\be \frac{2N_1N_2\epsilon\gamma_1^s}{(M^2-1)(M+\epsilon+s)(M+\epsilon-s)}.\ee
This vanishes when $\epsilon\to 0$, except when $s=M$. Taking this into account, the final result is indeed
\be\langle g\rangle =\frac{N_1N_2}{M}-\frac{N_1N_2M\gamma}{M^2-1}+\frac{N_1N_2\gamma^M}{M(M^2-1)}.\ee

\section{Conclusion}

By using a formulation in terms of matrix integrals, we developed a semiclassical approach to quantum conductance that is able to describe systems with tunnel barriers in the leads. Our result is in principle perturbative in the reflectivity, but exact in the number of channels, i.e. there is no large-$M$ expansion. However, the final result turns out to be polynomial in $\gamma$. In particular, we obtain a non-perturbative, exponentially small, contribution proportional to $\gamma^M$, not accessible to previous approaches which were restricted to leading orders in $1/M$.

Financial support from CAPES and from CNPq, grant 306765/2018-7, are gratefully acknowledged. We have profited from discussions with Jack Kuipers.


\begin{thebibliography}{99}

\bibitem{book1} M. Brack, R. Bhaduri, \emph{Semiclassical Physics} (CRC Press, 2018).

\bibitem{book2} M. C. Gutzwiller, \emph{Chaos in Classical and Quantum Mechanics} (Springer, 1990).

\bibitem{heller} E. J. Heller, \emph{The Semiclassical Way to Dynamics and Spectroscopy } (Princeton University Press, 2018).

\bibitem{berry} M. V. Berry, K. E. Mount. Rep. Prog. Phys. 35, 315 (1972).

\bibitem{book3} J. Ankerhold, \emph{Quantum Tunneling in Complex Systems: The Semiclassical Approach} (Springer, 2007).

\bibitem{book4} M. Razavy, \emph{Quantum Theory of Tunneling} (World Scientific, 2013).

\bibitem{holstein} B. R. Holstein, Am. J. Phys. 56, 338 (1988).

\bibitem{garg} A. Garg, Am. J. Phys. 68, 430 (2000).

\bibitem{hagedorn1} G. A. Hagedorn, A. Joye,  AMS IP Studies in Advanced Mathematics 16, 181 (2000).

\bibitem{hagedorn2} V. Gradinaru, G. A. Hagedorn, A. Joye,  J. Phys. A: Math. Theor. 43 474026 (2010).

\bibitem{toloza} J. H. Toloza,  Contemp. Math. 307, 299 (2002).

\bibitem{nazarov} Y. V. Nazarov, Y. M. Blanter, \emph{Quantum Transport: Introduction to Nanoscience} (Cambridge University Press, Cambridge, 2009).

\bibitem{qc1} R. A. Jalabert, H. U. Baranger and A. D. Stone,  Phys. Rev. Lett. 65, 2442 (1990).

\bibitem{qc2}  C. M. Marcus, A. J. Rimberg, R. M. Westervelt, P. F. Hopkins, A. C. Gossard,  Phys. Rev. Lett. 69, 506 (1992).

\bibitem{qc3} A. M. Chang, H. U. Baranger, L. N. Pfeiffer, K. W. West,  Phys. Rev. Lett. 73, 2111 (1994).

\bibitem{qc4} H. U. Baranger and P. A. Mello,  Europhys. Lett. 33, 465 (1996).

\bibitem{haake} F. Haake, \emph{Quantum Signatures of Chaos} (Springer, Berlin, 2010).

\bibitem{landauer} R. Landauer, IBM J. Res. Dev. 1, 223 (1957).

\bibitem{buttiker} M. B\"uttiker, Phys. Rev. Lett. 65, 2901 (1990).

\bibitem{bar1} S. Gustavsson, R. Leturcq, B. Simovi\v{c}, R. Schleser, T. Ihn, P. Studerus, K. Ensslin, D. C. Driscoll, and A. C. Gossard,  Phys. Rev. Lett. 96, 076605 (2006).

\bibitem{bar2} S. Hemmady, X. Zheng, T. M. Antonsen, Jr., E. Ott, and S. M.
Anlage,  Phys. Rev. E 71, 056215 (2005).

\bibitem{bar3} X. Zheng, S. Hemmady, T. M. Antonsen, Jr., S. M. Anlage, and
E. Ott, Phys. Rev. E 73, 046208 (2006).

\bibitem{bar4} U. Kuhl, M. Mart\'inez-Mares, R. A. M\'endez-S\'anchez, and H.-J.
St\"ockmann,  Phys. Rev. Lett. 94, 144101 (2005).

\bibitem{rmt} C. W. J. Beenakker,  Rev. Mod. Phys. 69, 731 (1997). 

\bibitem{BB2} P. W. Brouwer and C. W. J. Beenakker,  Phys. Rev. B 50, 11263 (1994).

\bibitem{BB} P. W. Brouwer and C. W. J. Beenakker,  J. Math. Phys. 37, 4904 (1996).

\bibitem{weak2} H. U. Baranger, R. A. Jalabert, A. D. Stone,  Phys. Rev. Lett. 70, 3876 (1993).

\bibitem{super1}  M. V. Berry, In: Asymptotics beyond All Orders. Edited by H. Segur et al. (Plenum Press, 1991).

\bibitem{super2} J. P. Boyd,  Acta Applicandae Mathematica 56, 1 (1999).

\bibitem{whitney} R. S. Whitney, Phys. Rev. B 75, 235404 (2007).

\bibitem{jacquod} D. Waltner, J. Kuipers, P. Jacquod, and K. Richter, Phys. Rev. B 85, 024302 (2012).

\bibitem{kuipersrichter} J. Kuipers and K. Richter, J. Phys. A: Math. Theor. 46 055101 (2013).

\bibitem{kuipers} J. Kuipers, J. Phys. A: Math. Theor. 42, 425101 (2009).

\bibitem{kanz1} P. Vidal and E. Kanzieper, Phys. Rev. Lett. 108, 206806 (2012).

\bibitem{vidal} P. Vidal,  J. Phys. A: Math. Theor. 48 265206 (2015).

\bibitem{kanz2} A. Jarosz, P. Vidal, and E. Kanzieper, Phys. Rev. B 91, 180203(R) (2015).

\bibitem{perez} S. Rodr\'iguez-Perez, R. Marino, M. Novaes, and P. Vivo, Phys. Rev. E 88, 052912 (2013).

\bibitem{Ehren1} İ. Adagideli, Phys. Rev. B 68, 233308 (2003).

\bibitem{Ehren2} S. Rahav, P. W. Brouwer, Phys. Rev. Lett. 95, 056806 (2005).


\bibitem{sieber1} M. Sieber K. Richter,  Phys. Scr. T 90, 128 (2001).

\bibitem{sieber2} K. Richter, M. Sieber, Phys. Rev. Lett. 89, 206801 (2002).

\bibitem{essen3} S. Heusler, S. Müller, P. Braun, F. Haake,  Phys. Rev. Lett. 96, 066804 (2006).

\bibitem{essen4} P. Braun, S. Heusler, S. Müller, F. Haake, J. Phys. A 39, L159 (2006).

\bibitem{essen5} S. Müller, S. Heusler, P. Braun, F. Haake, New J. Phys. 9, 12 (2007).

\bibitem{greg1} G. Berkolaiko, J. Kuipers,  Phys. Rev. E 85, 045201 (2012).

\bibitem{greg2} G. Berkolaiko, J. Kuipers,  J. Math. Phys. 54, 112103 (2013).

\bibitem{pedro} P. H. S. Bento, M. Novaes,  J. Phys. A: Math. Theor. 54, 125201 (2021).

\bibitem{matrix} M. Novaes,J. Phys. A 46, 502002 (2013).

\bibitem{trs} M. Novaes, Ann. Phys. 361, 51 (2015).

\bibitem{corr} M. Novaes,  J. Math. Phys. 57, 122105 (2016).

\bibitem{sebas} S. M\"uller, M. Novaes, Phys. Rev. E 98, 052208 (2018).

\bibitem{jacob} A. M. Mathai, \emph{Jacobians of Matrix Transformations and Functions o Matrix Arguments} (Singapore: World Scientific, 1997). 

\bibitem{selberg} P. J. Forrester and S. O. Warnaar,  Bull. Am. Math. Soc. 45, 489 (2008).

\bibitem{macdonald} I. G. MacDonald, \emph{Symmetric Functions and Hall Polynomials} (Oxford University Press, 1998).

\bibitem{kaneko} J. Kaneko, SIAM J. Math. Anal. 24, 1086 (1993).

\bibitem{kadell} K. W. J. Kadell, Adv. Math. 130, 33 (1997).



\end{thebibliography}
\end{document}